\documentclass[12pt,onecolumn,technote]{IEEEtran}

\usepackage{mathrsfs}
\usepackage{txfonts}
\usepackage [dvips] {graphicx}

\renewcommand\baselinestretch{2.0}

\ifCLASSINFOpdf

\else

\fi



\begin{document}

\title{Total Variation Minimization Based Compressive Wideband Spectrum Sensing for Cognitive Radios}

\author{Yipeng~Liu~\
        and~Qun~Wan


\thanks{Yipeng Liu and Qun Wan are with the Electronic Engineering Department,
University of Electronic Science and Technology of China, Chengdu,
611731, China, e-mail: ($ \{ liuyipeng,~wanqun\} $@uestc.edu.cn).}
\thanks{Manuscript received Month Day, 2011; revised Month Day, Year.}}

\markboth{Journal Title,~Vol.~X, No.~X, Month~Year}%
{Shell \MakeLowercase{\textit{et al.}}: Bare Demo of IEEEtran.cls for Journals}

\maketitle

\begin{abstract}

{\normalsize}
Wideband spectrum sensing is a critical component of a functioning
cognitive radio system. Its major challenge is the too high sampling
rate requirement. Compressive sensing (CS) promises to be able to
deal with it. Nearly all the current CS based compressive wideband
spectrum sensing methods exploit only the frequency sparsity to
perform. Motivated by the achievement of a fast and robust detection
of the wideband spectrum change, total variation minimization is
incorporated to exploit the temporal and frequency structure
information to enhance the sparse level. As a sparser vector is
obtained, the spectrum sensing period would be shorten and sensing
accuracy would be enhanced. Both theoretical evaluation and
numerical experiments can demonstrate the performance improvement.

\end{abstract}


\IEEEpeerreviewmaketitle

\section{Introduction}

\IEEEPARstart{W}{indband} spectrum sensing is a critical component
of a functioning cognitive radio system
\cite{sahai_spectrum_sensing}. To simultaneously grab any available
channel for secondary systems while avoiding harmful interference to
primary receivers, cognitive radios should be able to sense a wide
range of frequencies to find "blank spaces" in the licensed bands
which are not being used by primary users at a particular place and
time \cite{haykin_cognitive_radio}.

There are three ways to perform the wideband spectrum sensing: One
approach utilizes a bank of fixed narrowband bandpass filters. But
it requires an enormous number of Radio Frequency (RF) components
and bandpass filters, which lead to high cost. Besides, the number
of the bands is fixed and the filter range is always preset. Thus
the filter bank way is not flexible. Another approach use a tunable
bandpass filter (BPF). Based on a one-by-one search strategy, a BPF
can be employed to search one frequency band at a time. However,
this approach results in a significantly long search time
\cite{ghasemi_spectrum_sensing}. The third one is to implement a
wideband circuit using a single RF chain followed by high-speed DSP
to process the full range of the available spectrum at once
\cite{sahai_spectrum_sensing}. And the major challenge is the too
high sampling rate requirement.

Compressive sensing (CS) promises to be able to deal with the high
sampling rate problem \cite{donoho_compressed_sensing}
\cite{candes_introduction_CS}
\cite{candes_robust_uncertainty_principles}. To exploit the
frequency sparsity of the received wideband signal, CS is a novel
sampling technique that can sample sparse signals with sub-Nyquist
rates and performs accurate recovery. The survey shows that the
fixed spectrum allocation leads to a low spectrum utilization ratio
by the primary radios \cite{fcc_report}, which is also the reason
that cognitive radio can use the "blank spectrum holes" to settle
the spectrum scarcity problem. Furthermore, the low spectrum
utilization also says that the wiedeband signal is sparse in a wide
range of frequency domain with large probability.

CS has turned out to be an effective method to address the wideband
spectrum sensing. A pioneering work was done in
\cite{tian_compressed_wideband_sensing} \cite{tian_ditributed_cwss1}
\cite{tian_ditributed_cwss2}, where CS is exploited for wideband
spectrum sensing and the standard l1 norm minimization is used to
estimate the original spectrum. Other different types of
implementation structures for compressive wideband spectrum sensing
have been proposed in \cite{elsner_cwss}
\cite{wang_distributed_cwss} \cite{polo_cwss} \cite{liu_robust_cwss}
\cite{liu_fixed_cwss}. However, nearly all these methods exploit
only the frequency sparsity to perform compressive wideband spectrum
sensing.

In this paper, the temporal and frequency structure information is
jointly taken to incorporate in the compressive wideband spectrum
sensing via total variation minimization technique. In the wide
monitoring frequency band, though the nonzero frequencies are sparse
globally, yet in some local subbands, the nonzero entries may be
distributed densely \cite{fcc_report}. Therefore, a differential
operation on the power spectrum state vector would result in a
sparser one. Besides, It is typically that the spectrum occupancy
states change slowly in time \cite{fcc_report}. And when the initial
spectrum states is obtained, it is not the spectrum state but the
change of the spectrum states that controls the cognitive radios to
perform the dynamical secondary spectrum access. The gradient
operation on the spectrum state vectors at two adjacent detection
periods can give the the spectrum state change vector which can give
sufficient information to enable dynamical spectrum access.
Furthermore, as the primary signal is slow-varying, the spectrum
state change vector would be sparser than the spectrum state vector.
To achieve the differential operation in the joint temporal and
frequency domain, total variation \cite{romberg_tvm}, which can be
interpreted as the l1 norm of the (appropriately discretized)
gradient is minimized here. As the sparser vector is used, the
spectrum sensing performance would be enhanced.

In the rest of the paper, section II gives the sparse spectrum
signal model; In section III, the previous standard compressive
wideband spectrum recovery method is described; section IV develops
the total variation minimization based compressive wideband spectrum
sensing scheme; Section V briefly compares the performance; In
section VI, numerical experiments is taken to demonstrate the
performance enhancement of the proposed method; Finally Section VI
draws the conclusion.

\section{Signal Model}

Suppose that there exists several primary users within a wide
frequency range. The entire frequency band is segmented into
non-overlapping subbands. Primary users employ Frequency Division
Multiplexing with fixed channelization. The locations of these
subbands are assumed to be preset and known. But the corresponding
power spectral density (PSD) levels are slowly varying in different
spectrum sensing periods.

Different types of implementation structures for random sampling for
compressive wideband spectrum sensing have been proposed in
\cite{laska_aic1} \cite{laska_aic2} \cite{yu_mixed-signal}
\cite{mishali_xamping}. The operation to obtain the random samples
is usually called analog-to-information convertor (AIC). Briefly an
AIC may be conceptually viewed as an ADC operating at Nyquist rate,
followed by compressive sampling. Here the AIC is taken to sample
the analog baseband signal x(t). For illustration convenience, we
detail the algorithm in discrete setting as it did in
\cite{tian_compressed_wideband_sensing} \cite{tian_ditributed_cwss1}
\cite{tian_ditributed_cwss2} \cite{elsner_cwss}
\cite{wang_distributed_cwss} \cite{polo_cwss} \cite{liu_robust_cwss}
\cite{liu_fixed_cwss}. Denote the $ N \times 1 $ stacked vector at
the output of the ADC by

\begin{equation}
\label{eq2_1_ADC_output} {\bf{x}}_t  = \left[ {\begin{array}{*{20}c}
   {x_{tN} } & {x_{tN + 1} } &  \cdots  & {x_{\left( {t + 1} \right)N - 1} }  \\
\end{array}} \right]^T ,~~t = 0,~1,~ \cdots ~,~T-1
\end{equation}
where the time window for the ADC is set to be the Nyquist sampling
duration to recover the signal without aliasing. The corresponding
random samples of AIC is:

\begin{equation}
\label{eq2_2_AIC_output} {\bf{y}}_t  = \left[ {\begin{array}{*{20}c}
   {y_{tM} } & {y_{tM + 1} } &  \cdots  & {y_{\left( {t + 1} \right)M - 1} }  \\
\end{array}} \right]^T ,~~t = 0,~1,~ \cdots ~,~T-1
\end{equation}
With the $ M \times N $ random measurement matrix $ {\bf{\Phi }} $,
we can formulate the random sampling as:

\begin{equation}
\label{eq2_3_random_sampling}{\bf{y}}_t  = {\bf{\Phi x}}_t
\end{equation}

As only PSD is needed to be estimated here, the phase information of
the samples are not necessary. The PSD vector can be obtained
directly from the autocorrelation of the samples. Here the signal is
assumed to be zero-mean, wide-sense stationary. The autocorrelation
at lag \emph{j} are $ r_x (j) = E\left[ {x_n x_{n - j}^* } \right] $
and $ r_y (j) = E\left[ {y_n y_{n - j}^* } \right] $. The
autocorrelation vectors of the Nyquist samples
(\ref{eq2_1_ADC_output}) and the compressive samples
(\ref{eq2_2_AIC_output}) are

\begin{equation}
\label{eq2_4_ADC_autocrrelation} {\bf{r}}_{x,t}  = \left[
{\begin{array}{*{20}c}
   0 & {r_{x,t} ( - N + 1)} &  \cdots  & {r_{x,t} (0)} &  \cdots  & {r_{x,t} (N - 1)}  \\
\end{array}} \right]^T
\end{equation}
and

\begin{equation}
\label{eq2_5_AIC_autocrrelation} {\bf{r}}_{y,t}  = \left[
{\begin{array}{*{20}c}
   0 & {r_{y,t} ( - M + 1)} &  \cdots  & {r_{y,t} (0)} &  \cdots  & {r_{y,t} (M - 1)}  \\
\end{array}} \right]^T
\end{equation}
respectively. In practice, estimates of the autocorrelation are
obtained by averaging over several signal segments. Here it can be
easily get:

\begin{equation}
\label{eq2_6_Ry_Rx} {\bf{R}}_{\bf{y}}  = E\left[ {{\bf{y}}_k
{\bf{y}}_k^H } \right] = {\bf{\Phi R}}_{\bf{x}} {\bf{\Phi }}^H
\end{equation}
where

\begin{equation}
\label{eq2_7_Rx_element} \left[ {{\bf{R}}_{\bf{x}} } \right]_{i,j} =
r_x (i - j) = r_x^* (j - i)
\end{equation}
and

\begin{equation}
\label{eq2_8_Ry_element} \left[ {{\bf{R}}_{\bf{y}} } \right]_{i,j} =
r_y (i - j) = r_y^* (j - i)
\end{equation}
Then the relation between the Nyquist autocorrelation vector (NAV)
(\ref{eq2_4_ADC_autocrrelation}) and the compressive autocorrelation
vector (CAV) (\ref{eq2_5_AIC_autocrrelation}) can be formulated
\cite{polo_cwss} :

\begin{equation}
\label{eq2_9_ry_rx} {\bf{r}}_{\bf{y}}  = {\bf{Ar}}_{\bf{x}}
\end{equation}
where

\begin{equation}
\label{eq2_10_Dictionary} {\bf{A}} = \left[ {\begin{array}{*{20}c}
   {{\bf{\bar \Phi \Phi }}_1 } & {{\bf{\bar \Phi \Phi }}_2 }  \\
   {{\bf{\Phi \Phi }}_3 } & {{\bf{\Phi \Phi }}_4 }  \\
\end{array}} \right]
\end{equation}
Denoting the (i, j)-th element of $ {\bf{\Phi }}_{i,j} $ by $ \phi
_{i,j}^* $. The $ M \times N $ matrix has it entry as:

\begin{equation}
\label{eq2_11_sub_Dictionary_0} \left[ {{\bf{\bar \Phi }}}
\right]_{i,j}  = \left\{ {\begin{array}{*{20}c}
   0 & {i = 1,j = 1, \cdots ,N}  \\
   {\phi _{M + 2 - i,j} } & {i \ne 1,j = 1, \cdots ,N}  \\
\end{array}} \right.
\end{equation}
And the $ N \times N $ matrices $ {\bf{\Phi }}_1 $, $ {\bf{\Phi }}_2
$, $ {\bf{\Phi }}_3 $ and $ {\bf{\Phi }}_4 $ are:

\begin{equation}
\label{eq2_12_sub_Dictionary_1} {\bf{\Phi }}_1  = hankel\left(
{\left[ {{\bf{0}}_{N \times 1} } \right],\left[
{\begin{array}{*{20}c}
   0 & {\phi _{1,1}^* } &  \cdots  & {\phi _{1,N - 1}^* }  \\
\end{array}} \right]} \right)
\end{equation}

\begin{equation}
\label{eq2_13_sub_Dictionary_2} {\bf{\Phi }}_2  = hankel\left(
{\left[ {\begin{array}{*{20}c}
   {\phi _{1,1}^* } &  \cdots  & {\phi _{1,N}^* }  \\
\end{array}} \right],\left[ {\begin{array}{*{20}c}
   {\phi _{1,N}^* } & {{\bf{0}}_{1 \times (N - 1)} }  \\
\end{array}} \right]} \right)
\end{equation}

\begin{equation}
\label{eq2_14_sub_Dictionary_3} {\bf{\Phi }}_3  = toeplitz\left(
{\left[ {{\bf{0}}_{N \times 1} } \right],\left[
{\begin{array}{*{20}c}
   0 & {\phi _{1,N} } &  \cdots  & {\phi _{1,2} }  \\
\end{array}} \right]} \right)
\end{equation}

\begin{equation}
\label{eq2_15_sub_Dictionary_4} {\bf{\Phi }}_4  = teoplitz\left(
{\left[ {\begin{array}{*{20}c}
   {\phi _{1,1} } &  \cdots  & {\phi _{1,N} }  \\
\end{array}} \right],\left[ {\begin{array}{*{20}c}
   {\phi _{1,1} } & {{\bf{0}}_{1 \times (N - 1)} }  \\
\end{array}} \right]} \right)
\end{equation}
where \emph{hankel}(\textbf{a}, \textbf{b}) is a hankel matrix
(i.e., symmetric and constant across the anti-diagonals) whose first
column is \textbf{a} and whose last row is \textbf{b}; and
\emph{toeplitz}(\textbf{a}, \textbf{b}) is a toeplitz matrix (i.e.,
symmetric and constant across the diagonals) whose first column is
\textbf{a} and whose first row is \textbf{b}.

To monitor a broad band, high sampling rate is needed. It is often
very expensive. Besides, too many sampling measurements inevitably
increase the storage space and computational burden for (DSP), while
spectrum sensing requires a fast and accurate algorithm. CS provides
an alternative to the well-known Shannon sampling theory. CS is a
framework performing non-adaptive measurement of the informative
part of the signal directly on condition that the signal is sparse.
According to the FCC report \cite{fcc_report}, the allocated
spectrum is in a very low utilization ratio. That means the sparsity
inherently exists in the wideband spectrum. As $ \textbf{x}_t $ has
a sparse representation in frequency domain:

\begin{equation}
\label{eq2_16_Fourier_representation} {\bf{r}}_{\bf{x}}  = {\bf{\Psi
p}}\end{equation}
where $ {\bf{\Psi }} $ is the $ 2N \times 2N $
inverse discrete Fourier transform (DFT) matrix, and \textbf{p} is
the $ 2N \times 1 $ PSD vector. And Most of the entries of
\textbf{p} are zero or almost zero. When the number of nonzero
elements of \textbf{p} is \emph{S} ($ S \ll N  $), the signal is
said to be \emph{S}-sparse.

Here an $ M \times N $ random projection matrix $ {\bf{\Phi }} =
{\bf{S}} $ is used to sample signals, i.e. $ {\bf{y}}_t = {\bf{S}}
{\bf{x}}_t $, where $ M < N $ and $ \textbf{S} $ is a non-Uniform
Subsampling or Random Subsampling matrix which are generated by
choosing M separate rows uniformly at random from the unit matrix $
\textbf{I}_N  $. Combining (\ref{eq2_9_ry_rx}) and
(\ref{eq2_16_Fourier_representation}), the measurement samples can
be represented as

\begin{equation}
\label{eq2_17_autocorrelation_sparse_representation} {\bf{r}}_{y,t}
= {\bf{A\Psi p}}_t  = {\bf{Dp}}_t
\end{equation}
where \textbf{D} is
the $ 2M \times 2N $ dictionary for the compressive sparse
representation.

\section{The Standard Compressive Wideband Spectrum Sensing}

CS theory asserts that, if a signal has a sparse representation in a
certain space, one can use the random measurement to obtain the
samples and reconstruct it with overwhelming probability by
optimization on condition that the random measurement matrix
satisfies the restricted isometry property (RIP) which is a
condition on matrices ${\bf{\Phi }}$ which provides a guarantee on
the performance of ${\bf{\Phi }}$ in CS. It can be stated
as\cite{candes_robust_uncertainty_principles}
\cite{donoho_compressed_sensing}:

\begin{equation}
\label{eq3_1_RIP} \left( {1 - \delta _s } \right)\left\| {\bf{y}}
\right\|_2^2  \le \left\| {{\bf{\Phi y}}} \right\|_2^2  \le \left(
{1 + \delta _s } \right)\left\| {\bf{y}} \right\|_2^2
\end{equation}
or all the \emph{S}-sparse \textbf{y}. where $ \left\| {\bf{y}}
\right\|_2  = \left( {\sum\nolimits_i {\left| {y_i } \right|^2 } }
\right)^{{1 \mathord{\left/
 {\vphantom {1 2}} \right.
 \kern-\nulldelimiterspace} 2}} $   is the $ \mathscr{C}_2 $  norm of
the vector $ {\bf{y}} = \left[ {y_1 ,y_2 , \cdots ,y_M } \right]^T
$. The restricted isometry constant  $ \delta _s  \in \left( {0,1}
\right) $ is defined as the smallest constant for which this
property holds for all \emph{S}-sparse vectors \textbf{y}. There are
three kinds of frequently used measurement matrices, and the Random
Subsampling matrix \textbf{S} which is chosen here is one of the
three.

When the RIP holds, a series of recovering algorithm can reconstruct
the sparse signal. One is greedy algorithm, such as matched pursuit
(MP) \cite{mallet_MP}, orthogonal matched pursuit (OMP)
\cite{tropp_OMP}; another group is convex program, such as basis
pursuit (BP) \cite{chen_BP1} \cite{chen_BP2}, LASSO
\cite{tibshirani_lasso} \cite{efron_lasso} \cite{kim_lasso1}
\cite{kim_lasso2} and Dantzig Selector (DS) \cite{candes_DS}.
Comparing these algorithms, DS has almost the same performance as
LASSO. Both of these groups has the advantages and disadvantages.
Briefly it can be summarized that convex program algorithm has a
more reconstruction accuracy while greedy algorithm has less
computing complex.

To find the unoccupied spectrum for the secondary access, the signal
in the monitoring band is down-converted to baseband and sampling
the resulting analogue signal through an AIC that produces samples
at a rate below the Nyquist rate. Afterwards, the autocorrelation of
the samples is acquired.

Now the PSD of y(t) is estimated from the autocorrelation. With the
random measurement matrix $ {\bf{\Phi }} = {\bf{S}} $ and the random
measurement vector $ {\bf{y}}_t $ $ t = 0,~1,~ \cdots ~,~T $ , the
dictionary \textbf{D} can be the autocorrelation $ {\bf{r}}_{y,t} $
and the CWSS can be formulated as

\begin{equation}
\label{eq3_2_l0}\begin{array}{c}
 {\bf{p}} = \mathop {\arg \min }\limits_{\bf{p}} \left\| {\bf{p}} \right\|_0  \\
 {\rm{s}}{\rm{.t}}{\rm{.  }}{\bf{y}}_t  = {\bf{Dp}} \\
 \end{array}
\end{equation}
where $ \left\| {\bf{p}} \right\|_0 $  is the $ \mathscr{C}_0 $ norm
which counts the number of nonzero entries of the vector \textbf{p}.
The minimization of the $ \mathscr{C}_0 $  norm is the best
candidate to encourage the sparse distribution. However, it is not a
convex function and the formulated CWSS (\ref{eq3_2_l0}) is not a
convex programming.

To convert the CWSS model to a convex programming, the $
\mathscr{C}_0 $  norm is relaxing to the $ \mathscr{C}_1 $  norm.
Further more, to enhance the noise suppression performance, LASSO is
taken to do the CWSS. It is a shrinkage and selection method for
linear regression. It minimizes the usual sum of squared errors,
with a bound on the sum of the absolute values of the coefficients.
To get more accuracy, we can reformulate the CWSS based on LASSO as

\begin{equation}
\label{eq3_3_lasso}\begin{array}{c}
 {\bf{p}}_{LASSO}  = \mathop {\arg \min }\limits_{\bf{p}} \left\| {\bf{p}} \right\|_1  \\
 {\rm{s}}{\rm{.t}}{\rm{.  }}\left\| {{\bf{y}}_t  - {\bf{Dp}}} \right\|_2  \le \mu  \\
 \end{array}
\end{equation}
where $ \mu $ is a parameter bounding the amount of noise in the
data, and $ \left\| {\bf{p}} \right\|_1  = \sum\nolimits_i {\left|
{p_i } \right|} $  is the $ \mathscr{C}_1 $  norm of the vector $
{\bf{p}} = \left[ {p_(1) ,p_(2) , \cdots ,p_(2N) } \right]^T $. This
problem is a second order cone program (SOCP) and can therefore be
solved efficiently using standard software packages. It finds the
smallest $ \mathscr{C}_1 $ norm of coefficients among all the
decompositions that the signal is decomposed into a superposition of
dictionary elements. It is a decomposition principle based on a true
global optimization. A number of convex optimization software, such
as cvx \cite{grant_cvx}, SeDuMi \cite{sturm_sedumi} and Yalmip
\cite{lofberg_yalmip} can be used to solve the problem.

\section{The Proposed Compressive Wideband Spectrum Sensing}

The standard sparse recovery method based CWSS only exploits the
sparsity in frequency domain. To further enhance the sensing
performance, other structure information can be incorporated. One is
that the nonzero frequencies are in clusters
\cite{tian_compressed_wideband_sensing}, and the PSD level in every
cluster is almost similar. The differential of the PSD sequence
would result in a sparser sequence with most of the nonzero elements
locating at the band boundaries. The other structure information is
that the similarity of two PSD vectors in adjacent sensing periods
is very high as the primary signal is often slow-varying. The
differential of the adjacent PSD vector would also result in a
sparser vector than the original PSD vectors.

To integrate the differential operations in both temporal and
frequency domain, the total variation, which can be interpreted as
the $ \mathscr{C}_1 $ norm of the gradient, can be minimized. In CS,
the number of measurements required to successfully recover signal
is in proportion to the sparsity degree
\cite{candes_introduction_CS}. Hence, the differential operation can
reduce the number of measurements required, especially when there
are wideband signals in the sensing range. Combining sparsity of PSD
and differential PSD, we can obtain an total variation minimization
(TVM) based CWSS.

\subsection{The TVM Based Compressive Wideband Spectrum Sensing}

Assume for each spectrum sensing period, there are T measurement
vectors. After the autocorrelation operation, we formulate them into
a matrix and get that:

\begin{equation}
\label{eq4_1_Ry}\ {\bf{R}}_{2M \times T}  = \left[
{\begin{array}{*{20}c}
   {{\bf{r}}_{y,1} } & {{\bf{r}}_{y,2} } &  \cdots  & {{\bf{r}}_{y,T} }  \\
\end{array}} \right]
\end{equation}

\begin{equation}
\label{eq4_2_p}\ {\bf{P}}_{2N \times T}  = \left[
{\begin{array}{*{20}c}
   {{\bf{p}}_{t,1} } & {{\bf{p}}_{t,2} } &  \cdots  & {{\bf{p}}_{t,T} }  \\
\end{array}} \right]
\end{equation}
Thus we can reformulate
(\ref{eq2_17_autocorrelation_sparse_representation}) in matrix as:

\begin{equation}
\label{eq4_3_matrix autocorrelation_sparse_representation}\
{\bf{R}}_{2M \times T}  = {\bf{D}}_{2M \times 2N} {\bf{P}}_{2N
\times T}
\end{equation}

Then we can let:

\begin{equation}
\label{eq4_4}\ {\bf{\bar r}}_{2TM \times 1}  = {\mathop{\rm
vec}\nolimits} \left( {{\bf{R}}_{2M \times T} } \right)
\end{equation}

\begin{equation}
\label{eq4_5}\ {\bf{\bar p}}_{2TN \times 1}  = {\mathop{\rm
vec}\nolimits} \left( {{\bf{P}}_{2N \times T} } \right)
\end{equation}

\begin{equation}
\label{eq4_6}\ {\bf{B}}_{2TM \times 2TN}  = {\bf{D}}_{2M \times 2N}
\otimes {\bf{I}}_{T \times T}
\end{equation}

Thus we can get:

\begin{equation}
\label{eq4_7}\ {\bf{\bar r}}_{2TM \times 1}  = {\bf{B}}_{2TM \times
2TN} {\bf{\bar p}}_{2TN \times 1}
\end{equation}

The total variation can be regarded as 2-dimensional differential
as:

\begin{equation}
\label{eq4_8}\ {\bf{PV}} = \sum\limits_{i,j} {\left(
\begin{array}{l}
 \left| {P(i,j) - P(i - 1,j)} \right| + \left| {P(i,j) - P(i,j - 1)} \right| \\
  + \left| {P(i,j) - P(i + 1,j)} \right| + \left| {P(i,j) - P(i,j + 1)} \right| \\
 \end{array} \right)}
\end{equation}

Here the \textbf{PV} can be reformulated as:

\begin{equation}
\label{eq4_9}\ {\bf{PV}} = \left\| {{\bf{V}}_{8TN \times 2TN}
{\bf{\bar p}}_{2TN \times 1} } \right\|_1
\end{equation}
where

\begin{equation}
\label{eq4_10}\ {\bf{V}}_{8TN \times 2TN}= \left[
{\begin{array}{*{20}c}
   {\left( {{\bf{V}}_1 } \right)_{2TN \times 2TN} }  \\
   {\left( {{\bf{V}}_2 } \right)_{2TN \times 2TN} }  \\
   {\left( {{\bf{V}}_3 } \right)_{2TN \times 2TN} }  \\
   {\left( {{\bf{V}}_4 } \right)_{2TN \times 2TN} }  \\
\end{array}} \right]
\end{equation}

\begin{equation}
\label{eq4_11}\ {\bf{V}}_1  = \left[ {\begin{array}{*{20}c}
   1 & 0 & 0 & 0 &  \cdots  & 0 & 0 & 0 & 0  \\
   { - 1} & 1 & 0 & 0 &  \cdots  & 0 & 0 & 0 & 0  \\
   0 & { - 1} & 1 & 0 &  \cdots  & 0 & 0 & 0 & 0  \\
    \vdots  &  \vdots  &  \vdots  &  \vdots  &  \ddots  &  \vdots  &  \vdots  &  \vdots  &  \vdots   \\
   0 & 0 & 0 & 0 &  \cdots  & 0 & { - 1} & 1 & 0  \\
   0 & 0 & 0 & 0 &  \cdots  & 0 & 0 & { - 1} & 1  \\
\end{array}} \right]
\end{equation}

\begin{equation}
\label{eq4_12}\ {\bf{V}}_2  = \left[ {\begin{array}{*{20}c}
   { - 1_{1,1} } & 0 &  \cdots  & {1_{1,2N + 1} } & 0 &  \cdots  & 0  \\
   0 & { - 1_{2,2} } &  \cdots  & 0 & {1_{2,2N + 2} } &  \cdots  & 0  \\
    \vdots  &  \vdots  &  \ddots  &  \vdots  &  \vdots  &  \ddots  &  \vdots   \\
   0 & 0 &  \cdots  & { - 1_{2TN,2N + 1} } & 0 &  \cdots  & 1  \\
\end{array}} \right]
\end{equation}


\begin{equation}
\label{eq4_13}\ {\bf{V}}_3  = \left[ {\begin{array}{*{20}c}
   1 & { - 1} & 0 & 0 &  \cdots  & 0 & 0 & 0 & 0  \\
   0 & { - 1} & 1 & 0 &  \cdots  & 0 & 0 & 0 & 0  \\
    \vdots  &  \vdots  &  \vdots  &  \vdots  &  \ddots  &  \vdots  &  \vdots  &  \vdots  &  \vdots   \\
   0 & 0 & 0 & 0 &  \cdots  & 0 & 1 & { - 1} & 0  \\
   0 & 0 & 0 & 0 &  \cdots  & 0 & 0 & 1 & { - 1}  \\
   0 & 0 & 0 & 0 &  \cdots  & 0 & 0 & 0 & 1  \\
\end{array}} \right]
\end{equation}

\begin{equation}
\label{eq4_14}\ {\bf{V}}_4  = \left[ {\begin{array}{*{20}c}
   {1_{1,1} } & 0 &  \cdots  & { - 1_{1,2N + 1} } & 0 &  \cdots  & 0  \\
   0 & {1_{2,2} } &  \cdots  & 0 & { - 1_{2,2N + 2} } &  \cdots  & 0  \\
    \vdots  &  \vdots  &  \ddots  &  \vdots  &  \vdots  &  \ddots  &  \vdots   \\
   0 & 0 &  \cdots  & {1_{1,2N + 1} } & 0 &  \cdots  & { - 1_{2TN,2TN} }  \\
\end{array}} \right]
\end{equation}

%
%

Based on the above TVM formulation, the CWSS can be formulated as:

\begin{equation}
\label{eq4_15_TVM1}\
\begin{array}{c}
 {\bf{\bar p}}_{TVM1}  = \mathop {\arg \min }\limits_{{\bf{\bar p}}} \left\| {{\bf{V\bar p}}} \right\|_1  \\
 {\rm{s}}{\rm{.t}}{\rm{.  }}\left\| {{\bf{Y}} - {\bf{BP}}} \right\|_2  \le \mu  \\
 {\bf{\bar p}} = {\mathop{\rm vec}\nolimits} \left( {\bf{P}} \right) \\
 \end{array}
\end{equation}

In TVM based CWSS (TVM-CWSS) (\ref{eq4_15_TVM1}), a more sparser
vector is estimated. Thus the sensing performance would be enhanced.
%
%

There are T segments of measurement vectors in a sensing period.
However, when there are only one measurement vector at current time,
the TVM can exploit the sparsity in the differential between the
estimated PSD vector and the last time's PSD. Accordingly, with T =
2, the TVM-CWSS (\ref{eq4_15_TVM1}) would give the estimated
wideband spectrum at current time.

From multiple estimated PSD vectors, the spectrum holes can be
located. Also the starting time for of primary radios' vacancy can
be noticed. Experience tells us that for longer time that the
spectrum has been in vacancy, the spectrum would be more likely to
be unused.

\subsection{Performance Evaluation}

To Compare the TVM-CWSS and LASSO-CWSS, the required number for
successful recovery of the sparse signal is analyzed briefly. We
assume that there is no measurement noise or perturbation and let n
= 2N. With overwhelming probability, a single piece of PSD vector
can be successfully recovered for the convex programming for sparse
signal recovery on condition that \cite{candes_introduction_CS}:

\begin{equation}
\label{eq4_17_sample_number1}\ m \ge CS\log \left( {{n
\mathord{\left/
 {\vphantom {n S}} \right.
 \kern-\nulldelimiterspace} S}} \right)
\end{equation}
where C is some constant.

Suppose that the number of nonzero entries in each column of the
estimated PSD matrix \textbf{R} is S. Thus the total number of
measurement for successful LASSO-CWSS (\ref{eq3_3_lasso}) would be:

\begin{equation}
\label{eq4_18_sample_number2}\ m_{BP}  \ge TCS\log \left( {{n
\mathord{\left/
 {\vphantom {n S}} \right.
 \kern-\nulldelimiterspace} S}} \right)
\end{equation}

To analyze the performance of TVM-CWSS (\ref{eq4_15_TVM1}),we assume
that for each  the average cluster size is d, and the number of
clusters in a single PSD vector is K ($ K \ll S$). Thus we have S =
Kd. For a single PSD vector recovery, the TVM-CWSS reduce the number
of nonzero entries in the estimated sparse vector to be 2K.
Therefore, the number of measurement for a vector recovery via TVM
would be:

\begin{equation}
\label{eq4_19_sample_number3}\ m \ge KC\log \left( {{n
\mathord{\left/
 {\vphantom {n {\left( {2K} \right)}}} \right.
 \kern-\nulldelimiterspace} {\left( {2K} \right)}}} \right)
\end{equation}
Besides, As a slow-varying signal, only a small part of the
positions of the nonzero entries would changed. The variation
includes both the newly emerging nonzero entries and the disappeared
entries. Here we assume the number of variation in nonzero elements'
positions is $ \Delta $ ($ \Delta  \ll S $). Apart from the first
recovery of the first column when t = 0, the required measurement
number for each PSD column recovery is:

\begin{equation}
\label{eq4_20_sample_number4}\ m \ge \Delta C\log \left( {{n
\mathord{\left/
 {\vphantom {n {\left( \Delta  \right)}}} \right.
 \kern-\nulldelimiterspace} {\left( \Delta  \right)}}} \right)
\end{equation}

Thus, the total number of measurement for successful TVM-CWSS
(\ref{eq4_15_TVM1}) would be:
\begin{equation}
\label{eq4_20_sample_number5}
\ m_{TVM}  \ge C\left( {\left( {T - 1}
\right)\Delta \log \left( {{n \mathord{\left/
 {\vphantom {n {\left( \Delta  \right)}}} \right.
 \kern-\nulldelimiterspace} {\left( \Delta  \right)}}} \right) + K\log \left( {{n \mathord{\left/
 {\vphantom {n {\left( {2K} \right)}}} \right.
 \kern-\nulldelimiterspace} {\left( {2K} \right)}}} \right)} \right)
\end{equation}

As $ \Delta  \ll S $ and $ K \ll S$, the minimum number of
measurements for successful TVM-CWSS would be much less than
LASSO-CWSS'.

\section{Simulation Results}

\begin{figure}[!h]
 \centering
 \includegraphics[angle= 0, scale = 0.47]{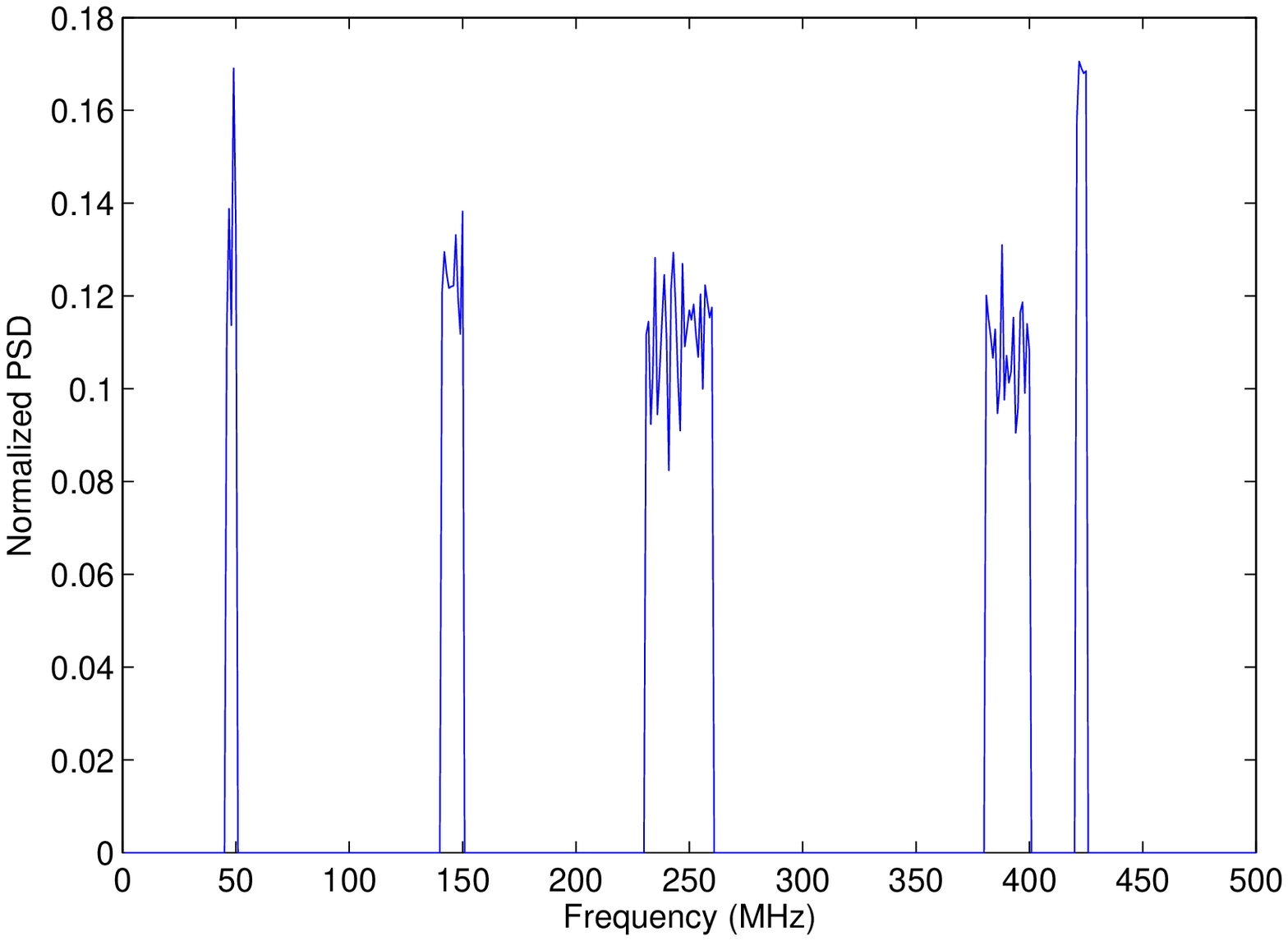}
 \caption{The normalized spectrum of noiseless active primary signals in the monitoring band.}
 \label{figure1}
\end{figure}

Numerical experiments are presented to illustrate performance of the
proposed TVM-CWSS for cognitive radio. Here we consider the initial
primary base band signal with its frequency range from 0Hz to 500MHz
as Fig. \ref{figure1} shows. The primary signal with random phase
are contaminated by a zero-mean additive Gaussian white noise (AGWN)
which makes the signal noise ratio (SNR) be 10dB. Five active block
dense bands are randomly located at eight subbands 46MHz - 50MHz,
56MHz - 60MHz, 141MHz - 150MHz,  161MHz - 170MHz, 231-260MHz, 381MHz
- 400MHz, 421MHz-425MHz, and 441MHz - 445MHz. The obtained wideband
signal has about 14.0$ \% $ nonzero entries, which is consistent to
the result of practical surveys. Normalizing the energy of the
active bands in the wideband,  One example of the PSD levels can be
given out as Fig. \ref{figure1} shows. Here we take the noisy signal
as the received signal \emph{x}(t). As CS theory suggests, we sample
\emph{x}(t) randomly at the sub-Nyquist sampling rate via AIC. The
resulted sub-sample vector is denoted as $ \textbf{y}_t $. Here only
one measurement vector are available at current time, and the
TVM-CWSS exploits the sparsity in the differential between the
estimated PSD vector and the last time＊s PSD.

\begin{figure}[!h]
 \centering
 \includegraphics[scale = 0.47]{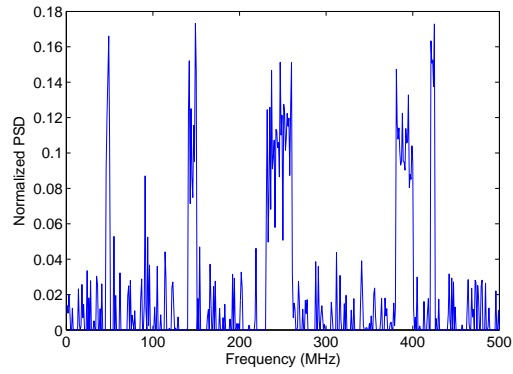}
 \caption{The normalized estimated spectrum via LASSO-CWSS in the monitoring band.}
 \label{figure2}
\end{figure}

\begin{figure}[!h]
 \centering
 \includegraphics[scale = 0.47]{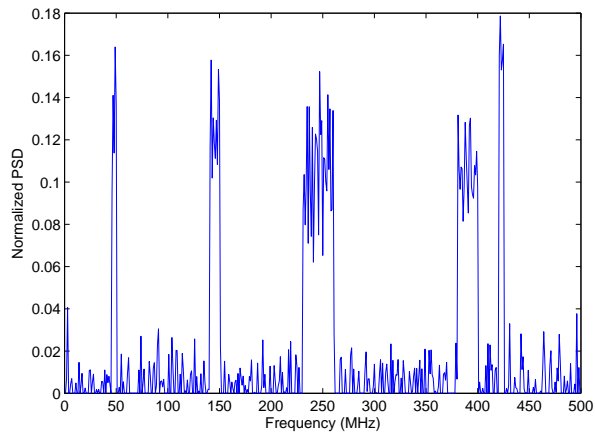}
 \caption{The normalized estimated spectrum via TVM-CWSS in the monitoring band.}
 \label{figure3}
\end{figure}

To make contrast, with the same number of samples and fixed nonzero
entries in the signal model, power spectrum with different methods
with 1000 tries averaged are given out in Fig. \ref{figure2} and
Fig. \ref{figure3}. The compressive measurement is obtained at $
0.25\% $  of the Nyquist rate. The parameter $ \mu $ is chosen to be
$ 0.05\left\| {{\bf{y}}_t } \right\|_2 $. They show that the
proposed TVM-CWSS gives better reconstruction performance. To make
details, it shows that there are too many fake spectrum points in
the subbands without active primary signals presented in Fig
\ref{figure2} for the standard LASSO. The noise levels of result
from LASSO-CWSS are quite high along the whole monitoring band; and
the real active frequencies can hardly distinguish themselves.
However, in Fig. \ref{figure3}, the active frequencies clearly show
up; the noise levels in the inactive bands are quite low. Besides,
at other sub-sampling rate, the proposed TVM-CWSS also outperforms
the standard LASSO-CWSS in the same way for wideband spectrum
sensing. It demonstrates that exploiting the differential frequency
sparsity enhances the sensing performance.

To further evaluate the performance, Monte Carlo simulations are
used to test the performance. The false alarm ratio vector can be
defined as:

\begin{equation}
\label{eq5_1_far} {\bf{p}}_F  = \left[ {\begin{array}{*{20}c}
   {p_{F,1} } & {p_{F,2} } &  \cdots  & {p_{F,Q} }  \\
\end{array}} \right]^T
\end{equation}
where Q is number of the divided subbands, and for the q-th subband,

\begin{equation}
\label{eq5_2_far1} p_{F,q}  = \frac{{TM\left( {H_{1,q} |H_{0,q} }
\right)}}{L}
\end{equation}
where L is the times of Monte Carlo Simulations and $ {TM\left(
{H_{1,q} |H_{0,q} } \right)} $ counts the times when the q-th
subband is in vain but the estimation gives the result that it is
occupied by primary radio. Similarly, the detection ratio vector can
be defined as:

\begin{equation}
\label{eq5_3_dr}\ {\bf{p}}_D  = \left[ {\begin{array}{*{20}c}
   {p_{D,1} } & {p_{D,2} } &  \cdots  & {p_{D,Q} }  \\
\end{array}} \right]^T
\end{equation}
where for the q-th subband,

\begin{equation}
\label{eq5_4_dr1} p_{D,q}  = \frac{{TM\left( {H_{1,q} |H_{1,q} }
\right)}}{L}
\end{equation}
where $ {TM\left( {H_{1,q} |H_{1,q} } \right)} $ counts the times
when the estimation correctly gives the result that the q-th sbuband
is occupied by primary radio.

\begin{figure}[!h]
 \centering
 \includegraphics[scale = 0.47]{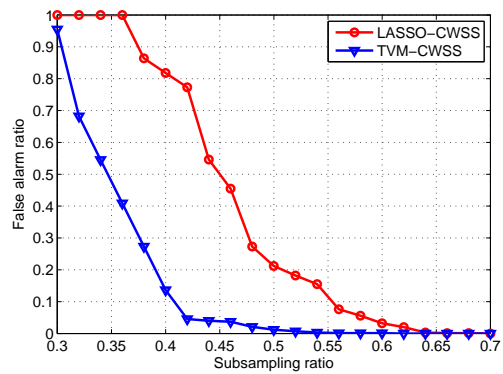}
 \caption{ The false alarm performance in the 5-th subband (150MHz - 230MHz) with different sub-sampling rate.}
 \label{figure4}
\end{figure}

\begin{figure}[!h]
 \centering
 \includegraphics[scale = 0.47]{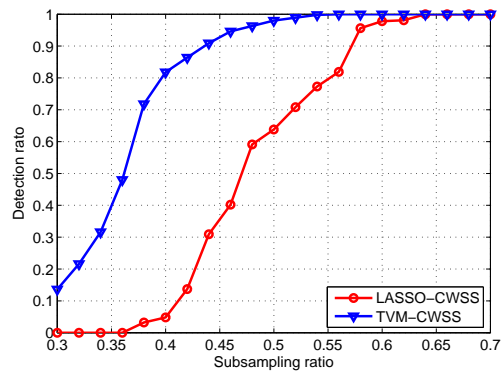}
 \caption{ The correct detection performance in the 6-th subband (231MHz - 260MHz) with different sub-sampling rate.}
 \label{figure5}
\end{figure}

For everyone of the L=1000 simuluations, five subbands are randomly
chosen from the eight ones to be active as it shows before. Fig.
\ref{figure4} illustrates the false alarm performance of the 5-th
subband (150MHz - 230MHz) via the measurement number for these two
CWSS, where the false alarm holds when at least one of the elements
in the 5-th subband is larger than one of the elements in the active
bands. And Fig. \ref{figure5} shows the detection ratio performance
of the 6-th subband (231MHz - 260MHz) via the measurement number,
where the detection holds when all elements in 6-th subband are
larger than all of the elements in the inactive bands. The
sub-sampling percentage to the Nyquist sampling is from 0.2 to 0.8.
Obviously we can see the TVM-CWSS (\ref{eq4_15_TVM1}) has a
remarkable performance gain against LASSO-CWSS (\ref{eq3_3_lasso})
for both the false alarm ratio and detection ratio performance.

\section{Conclusion}

This paper addresses the wideband spectrum sensing for cognitive
radio. Because the wideband primary signal often has the nonzero
frequencies in cluster, the frequency differential would result in a
sparser vector; and because the primary signal is often
slow-varying, the differential of the PSD vectors in adjacent
sensing period would also give out a sparser vector. Here the total
variation minimization is used to perform the joint
temporal-frequency differential for compressive wideband spectrum
sensing. Both performance analysis and stimulations demonstrate that
the TVM-CWSS has performance enhancement.

\section*{Acknowledgment}

This work was supported in part by the National Natural Science
Foundation of China under grant 60772146, the National High
Technology Research and Development Program of China (863 Program)
under grant 2008AA12Z306 and in part by Science Foundation of
Ministry of Education of China under grant 109139.

\ifCLASSOPTIONcaptionsoff
  \newpage
\fi

%
%
%
%
%

%
%

\end{document}